# Pilot-Wave Theory without Nonlocality

Paul Tappenden  paulpagetappenden@gmail.com
24 Place Castellane, 13006 Marseille, France
13th September 2022

> I saw the impossible done.
> Bell 1987: 160

> *a many-threads theory is ultimately just a hidden-variable theory where one simultaneously considers all physically possible worlds.*
> Barrett 1999: 184
> original emphasis

**Abstract**

It's generally taken to be established that no local hidden-variable theory is possible. That conclusion applies if our world is a *thread*, where a thread is a world where particles follow trajectories, as in Pilot-Wave theory. But if our world is taken to be a *set* of threads locality can be recovered. Our world can be described by a *many-threads* theory, as defined by Jeffrey Barrett in the opening quote. Particles don't follow trajectories because a particle in our world is a set of *elemental* particles following different trajectories, each in a thread. The "elements" of a superposition are construed as subsets in such a way that a particle in our world only has definite position if all its set-theoretic elements are at corresponding positions in each thread. Wavefunction becomes a 3D density distribution of particles' subset measures, the stuff of an electron's "probability cloud". Current Pilot-Wave theory provides a non-relativistic dynamics for the elemental particles (approximated by Many Interacting Worlds theory). EPR-Bell nonlocality doesn't apply because the relevant measurement outcomes in the absolute elsewhere of an observer are always in superposition.

**1 Pilot-Wave threads**

If our world is a pilot-wave world then hidden-variable particles follow trajectories, making our world a *thread* in Jeffrey Barrett's sense (Barrett 1999: 184). For any pilot-wave thread there's a range of possible trajectories consistent with the Born rule, determined by initial conditions, and if we consider the infinite set of all those possible threads then what we have in mind is a particular example of what Barrett calls a *many-threads* theory.[1]

Consider two pilot-wave threads whose measurable properties have been identical up to the performance of corresponding two-slit experiments where a single electron passes through each apparatus but they follow different trajectories which result in different measurement outcomes. Those threads can be said to be previously *parallel worlds* which *diverge*[2]. The complete set of pilot-wave threads can be thought of as initially parallel and subsequently partitioning into subsets which measurably differ. Since the set of possible particle trajectories for each thread must be determined by the Born rule to be empirically adequate it follows that the infinite set of initially parallel threads must partition in the same way as a hypothetical infinite set of empirically equivalent worlds where quantum measurement outcomes, of every variety, are determined stochastically.

That's to say, the corresponding divergent subset measures for the pilot-wave set and for the stochastic set must be equal. The pilot-wave set and the stochastic set have identical divergent structures, induced by measurement-like processes. Within a pilot-wave thread, measurement outcomes are fully determined but *subjective* probabilities can be assigned because of ignorance of the hidden trajectories. An equivalent stochastic theory posits *objective* probabilities and subjective probabilities then follow via the Principal Principle, the idea that observers should assign degrees of belief to the occurrence of a future event equal to what they believe is the objective probability of that occurrence.[3]

Two steps are now required to take us to a Pilot-Wave theory without nonlocality. Firstly, a change of perspective on the stochastic set which demonstrates how the probability of decay of an unstable particle can be understood as a rate of change of subset measure. This involves the hypothesis that a particle in an observer's environment is an infinite set of *elemental* particles, each in a thread. Secondly, the elimination of indeterminism by replacing the hypothetical stochastic set with the pilot-wave set and demonstrating that the resulting Many Pilot-Wave Worlds theory does *not* involve EPR-Bell nonlocality.

## 2 Situating the observer[4]

Since the advent of quantum mechanics it's been commonplace amongst physicists to think that we inhabit a single stochastic world in which quantum measurements always have single outcomes. Furthermore, given such a belief, it's at least plausible that there's such a thing as objective probability. It's *prima facie* plausible that the half-life of a francium-223 nucleus is a mind-independent property of that object which it has at any given moment. There can seem something elusive about the concept of propensity as a physical property but there's no killing objection to the hypothesis that objective probability of decay is a property of a francium-223 nucleus.

---

[1] As described by AntonyValentini (2010: 498-99).

[2] Following Simon Saunders (2010: 196), Alastair Wilson uses similar terms in a different context. Saunders' individual worlds don't contain hidden-variables (Wilson 2013, 2020).

[3] The term is from (Lewis 1980: 266). See (Tappenden 2021: §2.2) for further discussion.

[4] For some relevant discussion of this concept in the context of Pilot-Wave theory see Barrett 2021.

One can attempt to measure the half-life by taking a large sample of francium-223 nuclei and noting what *proportion* of them decay in a given period. By the Law of Large Numbers that yields some probability P *with a certain probability*. So, as is well know, a *measureable* frequency cannot be *identified* with a probability. However, by considering an *infinite* set of initially parallel stochastic worlds, each with a francium-223 nucleus at corresponding places, the measure of the subset of worlds in which decay has occurred after a certain period *must* be the probability for decay during that period, otherwise it would make no sense to assume the Law of Large Numbers when trying to measure probabilities.

To stress the point, the partitioning of a very large but finite set of the initially parallel stochastic worlds into subsets won't do for the coming argument. The measures of those subsets *cannot* be identified with the objective probability of decay of francium-223 for a given period. All that can be said is that the subset measures would have a high *probability* of being close to the objective probabilities. What's going to be required is the strict identification of the subset measures with the objective decay probabilities.

It's now time to follow a venerable tradition in physics and philosophy, which is to conduct a thought experiment. Imagine that Alice believes she inhabits a single stochastic world where the objective probability for the decay of a francium-223 nucleus is ½ per 22 minutes (0.0004 per second). Where might she be situated in an infinite set of parallel *stochastic* worlds? If she believes truly then she must inhabit one of them, in which case she has an infinite number of *Doppelgänger*, one in every other world but hers. However, Alice could be mistaken because an alternative interpretation of the setup is possible.[5]

It requires a hypothesis which turns out to be harmless despite being extremely counterintuitive. It's the hypothesis that there's a single Alice whose body is the set of *Doppelgänger*. That's to say, what Alice takes to be her body is an infinite set of isomorphic bodies. What Alice takes to be her world is an infinite set of worlds. Each individual stochastic world is a set-theoretic element of Alice's world so Alice's world can be said to be constituted by a set of *elemental* worlds. Likewise, the *Doppelgänger* in each elemental world are set-theoretic elements of Alice's body so they can be called elemental bodies. An electron in Alice's world is an infinite set of elemental electrons. In §4 I shall explain why set theory rather than mereology is being used here.

From Alice's point of view nothing has changed. She's a single subject who believes she has a body which is an individual, not an infinite set. She believes that she inhabits a single stochastic world. What's being called into question is the assumption that the relation between observers and *Doppelgänger* is one-to-one. All that follows is going to be built from the idea that that assumption is not needed and that abandoning it can advance our understanding of quantum theory.

---

[5] Introduced in (Tappenden 2017: §2) as the *Unitary Interpretation of Mind* and further developed in (Tappenden 2021).

Dropping an unnecessary assumption is generally taken to be a theoretical advance. In the light of that, the idea that an observer's world could be an infinite set of elemental worlds need not be seen as extravagant. If it's unnecessary to assume that individual concrete objects are not sets then set theory can be applied to physics in a novel fashion.

It would add an unnecessary complication to think of the elemental worlds as multiple regions in some sort of cosmological multiverse such as those described in (Tegmark 2007). That could well be relevant, but the key idea to keep in mind is that what we take to be individual objects in our environment can be construed as sets of objects. In §4 I shall say more about the intelligibility of applying set theory to physics in this way.

Returning to Alice. Beside her is a Geiger counter which is a set of elemental Geiger counters, each in a stochastic world. When Alice places a francium-223 nucleus in a chamber in her Geiger counter and starts a chronometer, all the *Doppelgänger* which are set-theoretic elements of her body move in concert in the parallel worlds, simultaneously transporting the nuclei to their corresponding places and starting elemental chronometers. Within $10^{-23}$ seconds, the light-time radius of a nucleus, the set of stochastic worlds has partitioned into a subset where decay has occurred and a subset where the nuclei remain intact. Necessarily, the measure of the decay subset is $10^{-23} \times 0.0004$.

After a few milliseconds the measure of the decay subset has increased $10^{20}$-fold and each Geiger counter in a decay world has been induced to emit an audible click. So Alice's Geiger counter *splits* in a fashion similar to that envisioned by Hugh Everett III (1957). It splits because it's a set of elemental Geiger counters which has partitioned into a subset which have emitted a click and a subset which haven't. The hypothesis is that one of the subsets is a non-elemental Geiger counter which has clicked because all of its elements have clicked. The other subset is a non-elemental Geiger counter which hasn't clicked because none of its elements have clicked. Alice's Geiger counter has split into two Geiger counters in definite click-states.

A little later the measure of decay worlds has further increased and Alice's body splits into a subset registering a click and a subset not registering a click. And shortly afterwards Alice splits into Alice$^1$ who hears a click by the time her chronometer has counted one second and Alice$^0$ who does not. Alice$^1$ is very surprised as she was expecting to wait for at least several minutes before hearing a click, whereas Alice$^0$ isn't at all surprised.

Alice$^1$'s world is a subset of Alice's world. Alice$^1$'s world is determined as being the set of elemental worlds in each of which is an element of her body, sustaining the cognitive state of hearing a click within one second of initiating the experiment. The measure of Alice$^1$'s world relative to Alice's world must be the probability that the elemental Geiger counters in each world will emit a click within one second, 0.0004. The measure of Alice$^0$'s world relative to Alice's world is 0.9996.

Everett's concept of splitting applies to a physical system which evolves into a superposition of *relative states*. So Alice's [Geiger counter + nucleus] system evolves into a superposition of [Geiger counter clicking + nucleus decayed] and [Geiger counter not clicking + nucleus undecayed]. On this interpretation of Everettian spitting, what physicists call the "elements" of a superposition are *subsets*. Post-measurement, there's one subset of the original set of stochastic worlds where the elemental worlds contain a decayed nucleus and a Geiger counter which has clicked and one subset where the elemental worlds contain an undecayed nucleus and a Geiger counter which hasn't clicked.

The phenomenon of Everettian splitting as interpreted here is very different from the splitting of amoeba to which Everett alluded, by way of being the closest analogy to hand (Barrett & Byrne 2012: 69). So I shall refer to it as *quantum fission*, following (Tappenden 2021 §2). It presents a problem which Everett didn't address. When Alice$^1$ refers to her Geiger counter she must believe that it was the Geiger counter of a moment earlier. Likewise Alice$^0$. But now there are *two different* Geiger counters (each of which is an infinite set). One has clicked and the other hasn't. How can it be that *both* of them *were* the original Geiger counter? A way to resolve that problem wasn't clearly articulated until 1996.

**2.1 Quantum fission and persistence**

*Stage theory* identifies objects with minimal temporal parts of their histories, objects are *stages* of their histories. The idea is presented in Ted Sider's seminal and nicely named article *All the World's a Stage*, though aspects of the idea had been suggested earlier (Sider 1996). Stage theory is tailor-made for quantum fission because it allows observers and objects to follow dendritic trajectories rather than just linear ones. It allows histories to be partially ordered series of stages. But that wasn't what Sider had in mind at the time. He later applied the idea to the problem of personal fission which was rendered particularly prominent by Derek Parfit. (Sider 2001: 201; Parfit 1984: 245-306).

Consider first of all how stage theory deals with the persistence of non-fissioning objects, such as an elemental Geiger counter in a stochastic world. The history of the object has momentary temporal parts each of which is a Geiger counter which bears *temporal counterpart* relations relative to all the other Geiger counters making up the history. So any given Geiger counter bears the temporal relation *was* to earlier Geiger counters in its history and *will be* to later Geiger counters. A persisting Geiger counter is one which bears temporal counterpart relations to other Geiger counters.

The counterpart concept is borrowed from modal theory where an object is taken to bear a relation to possible objects which it *might have been*. The idea that this wine glass might have been chipped is articulated as this wineglass bearing modal counterpart relations to various possible chipped wine glasses. Thus construed, the counterpart relation has a dimension of closeness linked to the concept of similarity. So the wine glass is in a sense modally closer to chipped wine glasses, more distant from smashed wine glasses and further still from wine glasses changed into vipers by wicked witches (or "quantum accidents"). But the concept of similarity isn't required for temporal counterparts because they bear relations of spatiotemporal continuity with each other. All the stages which are Geiger counters in a single history form a spatiotemporally continuous history and it's that which determines that they are temporal counterparts of each other.

Since the advent of Sider's stage theory there's been much discussion of the concept of stages in the philosophy of persistence. That discussion has been in the context of a debate between endurantists and perdurantists which has issued in what Thomas Pashby has called the *locational approach* to persistence. That this discussion has been divorced from stage theory is evident when Pashby writes:

> The basic question at the heart of the debate over persistence is this: how can the self-same thing be said to persist through time while its properties change?
> (Pashby 2016: 284)

Clearly stage theory is at odds with this statement since persistence is construed in terms of temporal counterpart relations between distinct objects. However, the locational approach has involved ideas about the concept of a stage which can be applied to stage theory as well as perdurantist *worm theory*, which identifies a persisting object with its world tube and takes the stages of that history to be temporal parts of the object. For stage theory, in contrast to worm theory, a persisting object is a temporal part of its history rather than *being* the history itself. The stages involved in both theories *are the very same things* which is, again, why a relation of similarity is not required between temporal counterparts any more than it between the stages of a history in worm theory.

Pashby writes:

> considerations from quantum theory suggest that the set of options considered thus far by the locational approach to persistence is ill-formulated or incomplete
> (Pashby 2016: 271)

Might applying the locational approach to stage theory help? A Geiger counter is to be thought of as a 3D object which is located at a particular region of space. Taking that into account helps to make clear what's involved in quantum fission construed in terms of an infinite set of stochastic worlds coupled with the unitary interpretation of mind. Alice's Geiger counter, at any given moment, is a set of elemental Geiger counters each located at an *elemental* spatial region. The spatial region occupied by Alice's Geiger counter is a spatial region which is a set of elemental spatial regions.

Alice's non-elemental Geiger counter fissions because it's a set of elemental Geiger counters which partitions when each stochastically evolves into a click-state or a non-click-state. So the spatial region where Alice's Geiger counter is located fissions into two distinct non-elemental spatial regions, one containing a non-elemental Geiger counter which has clicked and the other a non-elemental Geiger counter which hasn't clicked. To put it differently, Alice's Geiger counter evolves into a Geiger counter in an indefinite click-state because not all its elements are in the same click-state. That indefinite-state Geiger counter has two subsets which are Geiger counters in definite click-states because all their elements are in the same click-state. And those two Geiger counters occupy two distinct regions of non-elemental space which come to be occupied by the bodies of Alice$^1$ and Alice$^0$.

Alice's Geiger counter has *three distinct* Geiger counters as immediate future temporal counterparts, one in an indefinite click-state and two in definite click-states. According to stage theory each of the three *was* Alice's Geiger counter. But only the definite click states are observed, on the assumption that indefinite mental states don't exist.

Consider now applying the locational approach to the stage-theoretic analysis of a persisting object which is never located, such as an electron which hasn't interacted with anything since its creation in the Big Bang. The electron in Alice's world is an infinite set of electrons, each in an elemental world. And, for the maximal case, each elemental electron is at a different elemental location.

Now's the time to replace Alice's set of hypothetical stochastic elemental worlds with the empirically equivalent set of, supposedly real, pilot-wave worlds which includes all possible particle trajectories (and an infinite number of each if space isn't infinitely divisible). In each pilot-wave world a particle persists along its trajectory stage-theoretically by having, at any given moment, a temporal counterpart relation with all and only the other stages on the trajectory.

Necessarily, Alice's electron doesn't follow a trajectory since it's a set of elemental electrons following different trajectories. So how does Alice's electron stage-theoretically persist? Her electron is a set of elemental electrons, each with temporal counterpart relations with the other elemental electrons along its trajectory. The temporal counterparts of Alice's electron will be non-elemental electrons whose elements are the temporal counterparts of the elements of her electron. So the temporal counterparts of Alice's electron will never have a location, whilst electrons which are elements of Alice's electron always have elemental locations.

Stage theory is an established part of contemporary metaphysics, albeit somewhat neglected. Plausibly that's because it's so out of kilter with everyday semantics were we speak of a persisting object as being one-and-the-same thing from moment to moment, as we've seen in Pashby's characterisation of the question at the heart of persistence theory. Stage theory has bizarre consequences such as that there are many persons who existed in the last hour, each of whom you were. And you're not responsible for what you did yesterday. For legal purposes, it has to suffice that you were the person responsible. Such difficulties are a small price to pay for a coherent metaphysics for quantum mechanics.

The problem of persistence through quantum fission can be resolved via stage theory and, to stress the point, it yields an alternative concept of objects' histories. Histories are normally thought of as linear. The stages of a history are assumed to be well-ordered. Stage theory accommodates the possibility of dendritic histories where stages are partially-ordered. So it can be applied both to elemental worlds, where histories are linear, and to worlds which are partitioning sets of elemental worlds

Another problem for quantum fission theory is how to make sense of *subjective* probabilities. I've argued that the *objective* probabilities of Alice$^0$'s and Alice$^1$'s worlds relative to Alice's are 0.9996 and 0.0004. But a well-informed Alice is certain that both outcomes will occur which, following the Principal Principle in reverse, would seem to imply that she should assign an objective probability of 1 to each.

**2.2 Quantum fission and subjective probability**

When she believed that she inhabited a single stochastic world Alice believed that the objective probability of her nucleus decaying within the first second was 0.0004. On coming to adopt the alternative hypothesis about her situation in an infinite set of pilot-wave words this need not change, but now she also believes that she will be Alice$^1$ and she will be Alice$^0$. Again, it's being assumed that there's no third Alice in an indefinite state of belief. Whilst Alice's body will be a body having subsets which are bodies in two different cerebral states, registering a click or not, that's presumed to engender two definite cognitive states not a single indefinite cognitive state. We have no reason to believe that there are indefinite cognitive states such as simultaneously hearing and not hearing a click.

Notice also that Alice's non-elemental nucleus will be a nucleus which has decayed after one second and will be a nucleus which has not decayed. Thanks to stage theory that doesn't involve a contradiction, as it would if it were supposed that one-and-the-same nucleus does and does not decay. So the future occurrence of an event does *not* entail that the objective probability of its occurrence is 1 on this analysis of Alice's situation in an infinite set of pilot-wave worlds.

Following the Principal Principle, Alice assigns subjective probabilities, degrees of belief, to future events equal to what she believes the objective probabilities of those events to be. So she's certain that the superposition of relative states will occur since it's objective probability is 1. But she also knows that she won't be an observer hearing and not hearing a click. She'll split into Alice$^1$ hearing a click and Alice$^0$ not. So she assigns degrees of belief tracking the objective probabilities of those *definite* outcomes, 0.0004 and 0.9996 respectively. Alice is uncertain about what she'll observe because uncertainty is a cognitive state of assigning partial degrees of belief to multiple futures. We're in the habit of thinking of those futures as *alternatives* but there's no compelling reason to do so.

When it comes to Alice's initiating acts, Tim Maudlin poses a key question for quantum fission theorists. Suppose that Alice loves Erwin the cat and she's in the fiendish situation of having to put him either in Box A or Box B. She believes that there's a 2/3 chance of Erwin being killed in the former and a 1/3 chance in the latter.

> The squared amplitude of the branch in which Box A (or rather: the Box A-successor on that branch) is triggered is higher than the squared amplitude of the branch in which Box B is triggered. The ratio of the measures of the squared amplitudes distinguishes the branches from each other. The question is: Why, as an agent, should you care about this proportional difference in the squared amplitudes?
>
> (Maudlin 2019: 271)

Note that for stage theory the 'Box A-successor' is the box that Box A will be on that branch. Following the above analysis, the answer to Maudlin's question is simple: Alice should care about that proportional difference because it's the difference in the probability of there being a dead cat which was Erwin and the probability of there being a live cat which was Erwin. That's because the world which Alice inhabits is an infinite set of pilot-wave worlds which will partition into a subset of measures 2/3 which is a non-elemental world containing a dead cat that was Erwin, and a subset of measure 1/3 which is a world containing a live cat that was Erwin.

Alice's change of perspective need make no difference to the way she behaves, except, perhaps, in unusual scenarios[6]. When she believed that she inhabited a single stochastic world she did her best to increase the probability of the future she desired and that need not change when she comes to believe that her world is an infinite set of pilot-wave worlds and she'll fission into multiple futures.

I hope to have met Pilot-Wave theorists' secondary concerns about Many Worlds. Their prime concern has been Many Worlds theory's lack of particle trajectories, but that worry has clearly been met. The secondary concerns I take to be the trans-temporal identity and probability problems just discussed.

The application of set theory to physics, interpreting objects in our environment as infinite sets, is a radical move and I shall reflect on in it more after the next section. First of all, the task is to demonstrate that Many Pilot-Wave Worlds theory doesn't entail EPR-Bell nonlocality

## 3 Alice, Bob and Charlie

In Alice's world, an infinite set of pilot-wave worlds, particles don't follow trajectories because a particle in flight in Alice's world is a set of elemental particles following different trajectories. By hypothesis, Alice's particle could only follow a trajectory if all its set-theoretic elements were to follow corresponding trajectories in each pilot-wave world, which isn't the case for a many-threads theory. So the situation in Alice's world is very different for that in a single pilot-wave world where the trajectories of entangled particles are mutually determined at spacelike separation.

---

[6] For some discussion of unusual cases see (Tappenden 2011: §5).

However, there's no obvious way in which Alice would be able to know whether she inhabited a single pilot-wave world or a set of them since if she inhabited a single pilot-wave world it would be impossible to know *which* trajectory any given particle follows. The only reason Alice might choose to adopt the hypothesis that her world is a set of elemental worlds would be that in so doing she could recover locality with the possible advantage of helping her to develop a relativistic version of Pilot-Wave theory which might have as yet unforeseen empirical consequences. But to be assured that Alice's world indeed doesn't involve EPR-Bell nonlocality we need to consider the sorts of experiments which are now generally taken to show that no local hidden-variable theory is possible. The most straightforward way to do that is to consider the Greenberger-Horne-Zeilinger (GHZ) setup.[7]

At three spacelike separated locations spin measurements are made on three entangled particles relative to one of two axes which are randomly selected more quickly than the light-time between the locations. For four particular combinations of axis-selections out of the eight permutations, knowing any two of the outcomes makes it possible to predict the third with certainty. That's to say, the three pointer states are mutually determinate whilst a common cause is excluded by the rapid random selection of measurement bases.

The first thing to notice is that the conclusion requires measurement outcomes at spacelike separation to be definite relative to each other but that's *never* the case in Alice's world. What's in question here is the constitution of Alice's absolute elsewhere. A measurement system on Mars, such as a Geiger counter and a francium-223 nucleus, is an object in Alice's world. That entails that it's a set of measurement systems in all and only the elemental pilot-wave worlds which contain an element of Alice's body. If a measurement with multiple outcomes occurs on Mars the measurement system remains an object in Alice's world but it has subsets which are measurement systems showing different outcomes. In other words, immediately post-measurement, the measuring system on Mars is a *superposition of relative states* from Alice's point of view.

For the classic Alice and Bob setup involving a singlet state, if Alice makes her measurement she fissions into AliceUP and AliceDOWN, likewise Bob on Mars at spacelike separation. Relative to AliceUP and AliceDOWN Bob's in an indefinite state, and *vice versa*. So Alice's measurement doesn't induce Bob's measurement to be definite at spacelike separation, so EPR-Bell locality is preserved. However, AliceUP knows she'll see BobDOWN when light arrives from Mars and AliceDOWN knows she'll see BobUP. That's because Alice's and Bob's particles are entangled, and so their wavefunction is *nonseparable*. But nonseparability doesn't entail nonlocality which, to repeat, is a matter of whether in AliceUP's and AliceDOWN's worlds Bob's measurement at spacelike separation has a definite outcome. And *vice versa* for the Bobs.

Given that measurement-like processes are ubiquitous, macroscopic objects in Alice's absolute elsewhere will generally be in superposition. The Andromeda galaxy on Alice's simultaneity hyperplane will have myriad subsets that are galaxies which have evolved differently over the two million years which have elapsed since the light which she sees from there was emitted. Alice's distant absolute elsewhere is full of galactic superpositions.

---

[7] For a succinct account see (Maudlin 2019: 29-34).

The experiments which have actually been performed are less straightforward, requiring statistical evidence rather than a one-shot demonstration. However, that evidence is in the form of frequencies of correlations between definite measurement outcomes at spacelike separation. The assumption which underpins the conclusion that experimental tests of Bell's theorem confirm nonlocality is that measurement outcomes in our world are always definite, even when occurring in our absolute elsewhere. That's not the case for Many Pilot-Wave Worlds theory.

It needs to be stressed that the *apparent* nonlocality in elemental pilot-wave worlds *is not* a form of nonlocality. If Alice's world were a single pilot-wave world then there would be nonlocality in her world. But if Alice's world is a many-threads world where individual pilot-wave worlds are its set-theoretic *elements*, then there's no nonlocality in Alice's world. What's going on in the elemental pilot-wave worlds does not involve correlation between spacelike separated events in Alice's world because Alice's spacetime is a set of elemental spacetimes. Correlations between spacelike separated events in the set-theoretic *elements* of Alice's spacetime do not entail correlations between spacelike separated events in Alice's spacetime.

## 4 Resolution in set theory

However, the electron of a hydrogen atom in Alice's world is being construed as a *nonlocal object* in the following sense. Alice's spatial regions are sets of corresponding elemental spatial regions in pilot-wave worlds. For any of Alice's specific regions within her electron's "electron cloud" there will be a subset of electrons whose elements are located within regions which are set-theoretic elements of Alice's region. The measure of that subset is the absolute square of amplitude for the electron's wavefunction in Alice's region and is the probability that Alice would be an observer finding an electron in that region if she were able to look. The electron's wavefunction, the electron cloud, is interpreted as a 3D density distribution of its subset measures.

That's a radical idea. Objects in our environment are normally considered to be individuals, not sets. The dogma was challenged by Willard Van Orman Quine when he wrote:

> none of the utility of class theory is impaired by counting an individual, its unit class, the unit class of that unit class, and so on, as one and the same thing
>
> Quine1969: 31.[8]

---

[8] Quoted in (Tappenden 2017: 10).

Quine claimed that such set-theoretic individuals are "harmless", the reason being that the utility of set theory has been entirely to do with the construction of mathematical objects and that is unaffected by his proposal. But whilst harmless, Quine's idea might also be characterised as not very useful, in the sense that it has seemed to have no further utility than shepherding individuals into the set-theoretic fold. For that reason the idea has been neglected. I'm suggesting that Quinean individuals can at last be put to good use in physics by being identified with the set-theoretic elements of physical objects in an observer's environment. If we extend Quine's idea by going on to re-identify individuals as being any *set* of Quineian individuals that still does no harm to the utility of set theory in mathematics.[9]

It's commonly thought that sets differ from concrete objects in not having spatiotemporal extension but Quine's proposal demonstrates that that assumption isn't necessary. An apple which is a Quinean individual clearly has spatial extension, and temporal extension too according to worm theorists. Quine's crossing of the Rubicon on this issue opens the way for his proposal to be extended as I've suggested.

Now I can say something about why set theory should be preferred to mereology for providing a metaphysics for non-local objects and superposition. In (Tappenden 2000: 105-6) a mereological account of superposition was suggested, inspired by Michael Lockwood's articulation of the idea of a "superpositonal dimension" which he attributed to David Deutsch (Lockwood 1989: 232). The "elements" of a superposition where characterised as parts which were dubbed *superslices*. 'Super' from 'superposition' and 'slice' from 'timeslice' because superslices were somewhat akin to timeslices. The idea was that a superposition could be construed as an object which is "extended" in a non-spatiotemporal way. A particle in a superposition of spin on the z-axis, with a probability of 2/3 of being measured as *up* and 1/3 of being measured as *down*, might be compared with an apple which is green for 2/3 of its history and red for 1/3.

The problem remained as to what in quantum mechanics is to take the role analogous to durations for the apple? In what sort of non-spatiotemporal way might the superslices be extended? Set theory provides a candidate in the form of *subset measure*. Mereology is two-tier in the sense of positing wholes and parts whereas set theory is three-tier in that it posits sets, subsets and elements.

David Lewis attempted to join set theory and mereology at the hip by proposing his Main Thesis: "The parts of a class are all and only its subsets" (Lewis 1991: 7). But that's incompatible with Quine's proposal, as Lewis recognised. An elemental apple which is a self-membered singleton has parts such as peel and pips which are not its subsets. So, sad to say, Lewis's Main Thesis is not for us. Only a set theory where subsets are *not* parts of a set will do for the work in hand.

Construing sets of physical objects as themselves being physical objects can open up a Cantorian paradise for physics alongside that of mathematics because it allows us to use sets to provide a metaphysics for indefinite states, which are ontically challenging for Many Worlds theory. The idea is straightforward. We are free to construe a set of physical objects as having all and only the properties which its subsets share, with exceptions such as number of elements and value-definiteness. We are free to adapt set theory to the needs of physics in any way which looks promising, so long as it doesn't interfere with the use of set theory in mathematics.

---

[9] The term 'Quine atom' is sometimes used simply to denote a set which is its own sole element. I use the term 'Quineian individual' for objects characterized as in the quote.

The proposal may seem to be an absurd "category mistake" even though Quine has already demonstrated that there's no conceptual bar on construing an individual as being a set. Whence the air of absurdity? Perhaps because an alternative term for 'set' is 'class' and the notion of class brings to mind the idea of a category and surely a *category* can't be an individual. That *would* be a category mistake!

But the concept of a set is that of a collection rather than a category. A set of things can be ever so heterogeneous. We have no difficulty thinking of a heterogeneous collection of massive objects as being itself an object, the mereological sum, which has a mass equal to the sum of the masses of all its parts. Bizarre as it may seem, we can cope with thinking of the set of those very same massive objects as itself being an object which has the mass which its elements have in common; a mass which is thereby indefinite if all its elements don't have the *same* mass. Any set of massive objects in our environment would have indefinite position too.

So, a set of two apples in our environment, each of mass 100 grams, is an apple of mass 100 grams. That's on top of the 200 grams of mass for the mereological sum. So where's that further 100 grams of mass? If you were to throw the two you would feel the inertia of 200 grams not 300. What's going on? Time for another thought experiment.

Imagine that each of the apples were sent to one of two isomorphic planets. A pair of *Doppelgänger* on the two planets can be seen in the mind's eye to move isomorphically as if to each throw one of the apples. On the conventional interpretation each *Doppelgänger* is the body of a subject throwing an object of mass 100 grams. On the alternative interpretation required for Many Pilot-Wave Worlds theory, the set of the *Doppelgänger* is the body of a single subject throwing a single object which is the set of the two, mass 100 grams. The reason we can't feel the mass of the set of a pair of massive objects in our environment is that we cannot interact with the set. In order to do so one would have to be a subject whose body is a set of two *Doppelgänger* which are each able to interact isomorphically with one of the two objects. Accepting that strange idea opens the way to a Pilot-Wave theory without nonlocality. It has a wider consequence too.

**5 Dropping the pilot**

Many Interacting Worlds theory aims to replace the non-relativistic Pilot-Wave dynamics of hidden-variable particles with that of particle interactions (Hall *et. al.* 2014; Sebens 2015)[10]. As Charles Sebens writes:

> Unlike the many worlds of the many-worlds interpretation, these worlds are fundamental, not emergent; they are interacting, not causally isolated; and they never branch.
> Sebens 2015: 267

Specifically referring to a set of independently evolving Pilot-Wave worlds, Sebens argues that the wave-guided dynamics can be replaced by that of inter-world particle interactions obeying Newtonian force laws. The implication is that there may be something to be gained by dropping a fundamental ontic commitment to waves.

Many Interacting Worlds theory currently faces two fully-admitted difficulties. One is that the number of threads needs to be finite so it can only approximate standard non-relativistic quantum mechanics. The reason is that, because observers are situated in individual threads, probabilities are derived from self-location uncertainty[11]. That problem is resolved via the alternative perspective because objective probabilities can be identified with subset measures on *infinite* sets in the way that I've described and subjective probabilities can thereby be derived via the Principal Principle [12]. No appeal to self-location uncertainty is required.

The interacting worlds to which Sebens refers are the threads of a many-threads theory and Many Interacting Worlds theorists have always situated observers within threads. But if we adopt the alternative perspective, identifying observers' worlds with sets of threads, a similar structure to that of the Many Pilot-Wave Worlds theory is retrieved. The branches, which are subsets of threads, will emerge via measurement-like events and those subsets can constitute worlds for observers such as Alice$^1$ and Alice$^0$. Those worlds are indeed, for all practical purposes, causally isolated, as in

---

[10] See (Vaidman 2014: §10) for some discussion of this type of approach to quantum theory.

[11] Hall, Deckert and Wiseman write:

> The MIW [Many Interacting Worlds] approach can only become equivalent to standard quantum dynamics in the continuum limit, where the number of worlds becomes uncountably infinite. (2014 §1)

Sebens argues that finitude is called for because the intelligibility of a subject's assignment of probabilities to alternative futures is based on '*self-locating uncertainty*' which requires a 'basic indifference principle' 2015: 282, original emphasis).

[12] In (Tappenden 2021: 6390) it's suggested that a version of the Principle Principle might be justified via the Deutsch-Wallace argument.

Many Worlds theory. Again, a particle's wavefunction in Alice's world is interpreted as a 3D density distribution of that particle's subset measures (the subsets being what physicists call the "elements" of a superposition). The significant difference with Many Pilot-Wave Worlds as discussed above is that the evolution of the density distribution is driven by interactions between particles rather that the pilot-wave dynamics.

The other difficulty is that Many Interacting Worlds theory, like Pilot-Wave theory, is nonlocal. That problem is resolved in the same way as before if Alice's world is a set of interacting elemental worlds. It matters not whether the elemental worlds are of the pilot-wave or interacting variety. It will still be the case that measurements with multiple outcomes will *never* yield definite outcomes in Alice's absolute elsewhere, those outcomes will *always* be superpositions of relative states.

**6 Conclusion**

I've argued that a version of Pilot-Wave theory is available which doesn't entail EPR-Bell nonlocality. It combines Many Worlds and Pilot-Wave theories in a novel way which meets Pilot-Wave theorists' longstanding concerns about the former. The ideas can also yield a causally local Many Interacting Worlds theory which does not need to ground probability in self-location uncertainty and which can exactly model non-relativistic quantum mechanics because able employ an infinite set of interacting worlds. It's perhaps not unreasonable to hope that this discovery may be of help in developing relativistic hidden-variable theories.

**Acknowledgements**

My thanks to Simon Saunders for a careful reading of an earlier draft which revealed a crucial error. To Jeff Barrett and Chip Sebens for some helpful comments and to Aurélien Drezet for the benefits of an extended discussion following a talk I gave for the Grenoble philosophy of physics group. Thanks also to a reviewer who encouraged me to say more about the details of the metaphysics and indicated a number of issues in need of resolution.